\documentclass{article}
\usepackage{amsfonts}
\usepackage[utf8]{inputenc}
\usepackage{amsmath, amssymb, amsthm,  bm}
\usepackage{graphicx,url}
\usepackage[colorlinks=true]{hyperref}
\usepackage{tikz}
\usepackage{color}
\usepackage{xcolor}
\usepackage{tikz}
\usepackage[
backend=biber,
style=apa,
sorting=nyt
]{biblatex}
\DeclareSourcemap{
  \maps[datatype=bibtex]{
    \map{
      \step[fieldset=issn, null]
      \step[fieldset=doi, null]
      \step[fieldset=url, null]
      \step[fieldset=urldate, null]
    }
  }
}

\usetikzlibrary{shapes, patterns, decorations, fit, intersections}
\newtheorem{theorem}{Theorem}

\newtheorem{assumption}{Assumption}

\definecolor{bgColorInterpretation}{RGB}{240,240,240} 
\definecolor{bgColorTitleInterpretation}{RGB}{10,140,52} 
\tikzstyle{interpretationBox} = [draw=bgColorTitleInterpretation, fill=bgColorInterpretation, very thick, rectangle, rounded corners,
inner sep=10pt, inner ysep=20pt] 
\tikzstyle{interpretationBoxTitle} =[fill=bgColorTitleInterpretation,
text=white]
\providecommand{\abs}[1]{\lvert#1\rvert}
\providecommand{\norm}[1]{\lVert#1\rVert}

\def\uniset{{\rm 1\kern-.40em 1}}

    \numberwithin{equation}{section}
    
\def\bkR{{\rm I\kern-.17em R}}
    \theoremstyle{definition}

\def\citecomplete#1{\citeauthor{#1} (\citeyear{#1})}

\begin{document}

\title{\textbf{Estimating Separable Matching Models\thanks{%
The authors are  grateful to Cl\'ement Montes for superb research assistance, and to Antoine Jacquet for his detailed comments.}
}}
\author{Alfred Galichon\thanks{%
New York University and Sciences Po. Support from ERC grant EQUIPRICE No.
866274 is acknowledged.} \and Bernard Salani\'e\thanks{%
Columbia University.}}
\date{\today }
\maketitle

\begin{abstract}
In this paper we propose two simple methods to estimate models of matching
with transferable and separable utility introduced in~\citecomplete{cupid:restud}. The
first method is a minimum distance estimator that relies on the generalized
entropy of matching. The second relies on a reformulation of the more special but popular \citecomplete{choo-siow:06} model; it uses generalized linear
models (GLMs) with two-way fixed effects.
\end{abstract}


\section*{Introduction}

The estimation of models of two-sided matching has made considerable
progress in the past decade. While some of this work has used matching under
non-transferable utility, many applications have focused on markets where
utility is transferable. The pioneering contribution of~\citecomplete{choo-siow:06}
introduced a simple and highly tractable specification.  They used their model to estimate the effect of the 1973
liberalization of abortion in the US on marriage outcomes. In doing so, they
used a nonparametric estimator of the matching patterns. Their specification is a natural extension of the multinomial
logit model, and it has become quite popular.

The \citeauthor{choo-siow:06} specification rests on three
main assumptions that will be defined later in the paper: separability; large market; and standard type I extreme value
random utility.
In~\citecomplete{cupid:restud}, we showed that the third,
distributional assumption is not necessary: for any (separable) distribution
of the errors, the joint surplus is nonparametrically identified.  
The nonparametric estimator of \citeauthor{choo-siow:06} was
feasible in their case as they only conditioned on the ages of the partners in a couple.
It breaks down, however, when more covariates are considered as  matching cells become too small; and by construction, it does not allow for parameterized error distributions.
Structural models of
household behavior also naturally introduce parameters.

In all of these cases, the analyst must  resort to parametric models.  This note shows two
very simple methods to estimate parametric versions of separable matching
models with perfectly transferable utility, with special emphasis on the~%
\citeauthor{choo-siow:06} model and more generally on ``semilinear'' models, where the joint
surplus is linear in the parameters (again, to be formally defined later in the paper).

Our first method applies a minimum-distance estimator to the identification
equation derived in~\citecomplete{cupid:restud}, which relates the joint surplus to the
derivatives of a generalized entropy function evaluated at the observed
matching patterns. For any fixed distribution of the error terms, the
generalized entropy can be evaluated and differentiated, numerically if
needed. The estimator selects parameter values and also provides a simple
specification test. In semilinear models, the estimator can be obtained in
closed form.

The second method we present applies more specifically to the semilinear
Choo and Siow model. We show that the moment-matching estimator we described
in~\citecomplete{cupid:restud} can be reframed as a generalized linear model, more
specifically as the pseudo-maximum likelihood estimator of a Poisson
regression with two-sided fixed effects. This is available as \texttt{%
linear\_model} in the \texttt{scikit-learn} library in \texttt{Python}, as 
\texttt{fepois} in the \texttt{R} package \texttt{fixest} and as \texttt{%
ppmlhdfe} in Stata, among other common statistical packages.

We conclude with a brief discussion of the pros and cons of these two
methods. Both are coded in a Python package called 
\texttt{cupid\_matching} that is available on the standard repositories\footnote{%
See \url{http://bsalanie.github.io} for more information, and %
\url{https://share.streamlit.io/bsalanie/cupid_matching_st/main/cupid_streamlit.py}
for an interactive Streamlit app that demonstrates solving and estimating a 
\citecomplete{choo-siow:06} model.}.

\section{The Model}

This paper applies to a bipartite matching market with perfectly
transferable utility. For simplicity, we refer to potential partners as
``men'' and ``women''. We use the same notation as in~\citecomplete{cupid:restud}. We
assume that the analyst can only observe which of a finite set of \emph{%
types} each individual belongs to. Men and women of a given type differ along some dimensions that they all
observe, while the analyst does not.  Each man $i\in \mathcal{I}$ belongs to
one group of (observable) type $x_{i}\in \mathcal{X}$; and, similarly, each woman $j\in \mathcal{J%
}$ belongs to one (observable) type $y_{j}\in \mathcal{Y}$. We will say that ``man $i$
is of type $x$'' and ``woman $j$ is of type $y$.'' We denote $\mu_{ij}$
the indicator function for a matching between man $i$ and woman $j$, which
is equal to $1$ if $i$ and $j$ are matched and to $0$ otherwise. Similarly, $%
\mu_{i0}$ and $\mu_{0j}$ are the indicator of $i$ or $j$ to remain
unmatched, respectively. Without loss of generality, we assimilate $\mathcal{%
X}$ to $\{1,\ldots,X\}$ and $\mathcal{Y}$ to $\{1,\ldots,Y\}$. As $\mathcal X$ and $\mathcal Y$ will later serve as choice sets of partners types for women and men, respectively, and as the marital options needs to include remaining unmatched, we shall add the option to remain unmatched $0$ to these sets and denote   $\mathcal X_{0} = \mathcal X \cup \{0\}$ and $\mathcal Y_{0} = \mathcal Y \cup \{0\}$ the respective sets of marital options of women and men. 

We denote $n_{x}$ the mass of
men of type $x\in \mathcal X$, and $m_{y}$ the mass of women of type $y \in \mathcal Y$. We denote $\bm{q}=(\bm{n}, %
\bm{m})$ the vector that collects the margins $\bm{n} $ and $\bm{m}$ of the
problem

In addition to the margins~$\bm{q}$, the analyst observes matchings at the
type level. We denote $\mu_{xy}$ the mass of the couples where the man
belongs to type $x$, and where the woman belongs to type $y$, which is formally defined as $\mu_{xy} = \sum_{ i\in \mathcal{I}, j\in\mathcal{J}} \mu_{ij}1\{ x_i =
x\}1\{ y_j = y\}$. We also denote $\mu_{x0} = \sum_{ i\in \mathcal{I}} \mu_{i0}1\{ x_i = x\}$,
and $\mu_{0y} = \sum_{ j\in\mathcal{J}} \mu_{0j}1\{ y_j = y\}$ the mass of
single individuals who are respectively men of type $x$ and women of type $y$. 
We will be interested in the limiting market with a large number of men in
any type $x$, and of women in any type $y$. Since the
problem is homogeneous, we shall normalize  the total mass $N$ of households to one; that is, we rescale $\bm{\mu}$ and $\bm{q}$ by a multiplicative factor $N$ such that $ \sum_{x,y}\mu_{xy}+\sum_x \mu_{x0}+\sum_y \mu_{0y} =1$. Again, we use the boldface notation  $\bm{\mu}$ to denote the vector of matching numbers. We denote $\mathcal A = \mathcal X \times \mathcal Y \cup \mathcal X \times \{ 0 \} \cup \{0 \} \times \mathcal Y$ the set of possible marital arrangements (matched household of type $xy$, or single households of type $x0$ or $0y$), so that $\bm{\mu}$ is a vector of $\mathbb R^{\mathcal A}$.
 
A \emph{matching} is the specification of who matches with whom. It is \emph{%
feasible} if each individual is matched to at most one partner. It is \emph{%
stable} if no individual who has a partner would prefer to be single, and if
no two individuals would prefer forming a couple over their current
situation.

We model the 
\emph{joint surplus} $\tilde{\Phi}_{ij}$, which is the sum of the cardinal utilities
that both a man $i$ and a woman $j$ jointly obtain by being matched
together, and we assume a \emph{separable } matching surplus:

\begin{assumption}[Separability]
\label{ass:separ}  There exist a vector $\bm{\Phi}$ in $\mathrm{I\kern-.17em
R}^{X\times Y}$ and random terms $\bm{\varepsilon}$ and $\bm{\eta}$ such that

\begin{itemize}
\item[(i)] the joint utility from a match between a man $i$ of type $x\in 
\mathcal{X}$ and a woman $j$ of type $y\in \mathcal{Y}$ is 
\begin{equation}  \label{eq:sepPhi}
\tilde{\Phi}_{ij}=\Phi _{xy}+\varepsilon _{iy}+\eta _{xj},
\end{equation}

\item[(ii)] the utility of a single man $i$ is $\tilde{\Phi}%
_{i0}=\varepsilon_{i0}$,

\item[(iii)] the utility of a single woman $j$ is $\tilde{\Phi}%
_{0j}=\eta_{0j}$,
\end{itemize}

where, conditional on $x_{i}=x$, the random vector $\bm{\varepsilon}%
_{i}=(\varepsilon _{iy})_{y\in\mathcal{Y}_{0}}$ has probability distribution 
$\mathbb{P}_{x}$, and, conditional on $y_{j}=y$, the random vector $\bm{\eta}%
_{j}=(\eta _{xj})_{x\in\mathcal{X}_{0}}$ has probability distribution $%
\mathbb{Q}_{y}$. The distributions $\mathbb{P}_{x}$ and $\mathbb{Q}_{y}$
have full support and a density with respect to the Lebesgue measure. The
variables 
\begin{equation*}
\max_{y\in \mathcal{Y}_0} \; \abs{\varepsilon_{iy}} \; \mbox{ and } \;
\max_{x\in \mathcal{X}_0} \; \abs{\eta_{xj}}
\end{equation*}
have finite expectations under $\mathbb{P}_x$ and $\mathbb{Q}_y$
respectively.
\end{assumption}

Separability allows for a restricted form of ``matching on unobservables'';
it rules out interaction terms on characteristics that are unobserved on
both sides of the market, e.g.\ some unobserved preference of man $i$ for
some unobserved characteristics of woman $j$.

\citecomplete{csw:17} and \citecomplete{cupid:restud} showed that under separability, at any
stable matching $\bm{\mu}$ there exist two matrices $\bm{U}$ and $\bm{V}$
such that for all $(x,y)$, $U_{xy}+V_{xy} = \Phi_{xy}$, and $U_{x0}=V_{0y}=0$%
, and such that man $i$ of type $x$ is assigned option $y=0,1,\ldots,Y$
which maximizes $U_{xy}+\varepsilon_{iy}$ (where option $0$ means remaining
unmatched, and option $y\neq 0$ means being matched with a woman of type $y$);
similarly woman $j$ of type $y$ is assigned option $x=0,1,\ldots,X$ which
maximizes  $V_{xy}+\eta_{xj}$.

\subsection{Generalized Entropy}

Consider the classic ``Emax'' function $G_x$ defined as follows. In this paragraph we let $\bm{U}%
=(U_1,\ldots,U_{Y})$ be a $Y$-dimensional vector. Then we define 
\begin{equation*}
G_x(\bm{U}) = E_{\mathbb{P}_x}\max\left(\max_{y\in \mathcal{Y}}
(U_{y}+\varepsilon_{iy}),  \varepsilon_{i0}\right).  
\end{equation*}
As a maximum of linear functions, $G_x$ is a convex function. We denote $%
\partial G_x(\bm{U})$ its subgradient; because of the assumptions made on $%
\mathbb{P}_{x}$, it is a singleton almost everywhere.

Now take the Legendre-Fenchel transform of $G_x$: for any $(\nu_1,\ldots,
\nu_{Y})$ such that $\sum_{y\in \mathcal{Y} }\nu_y\leq 1$, we define 
\begin{equation*}
G^\ast_x(\bm{\nu}) = \max_{\bm{U}} \left(  \sum_{y\in \mathcal{Y}}
\nu_{y}U_y-G_x(\bm{U})  \right).  
\end{equation*}
It is another convex function; and since $G_x$ is convex, $G_x$ is the
Legendre-Fenchel transform of $G^\ast_x$. As a consequence,  
\begin{equation*}
\bm{\nu} \in \partial G_x(\bm{U})  \; \mbox{if and only if} \;  \bm{U} \in
\partial G^\ast_x(\bm{\nu}).  
\end{equation*}
This convex duality is at the core of the identification and inference
results in~\citecomplete{cupid:restud}.

Defining $H_y$ and $H^\ast_y$ in the same way, we get the \emph{generalized
entropy}: for any feasible matching $\bm{\mu}$, 
\begin{equation}  \label{eq:defE}
\mathcal{E}(\bm{\mu},\bm{q}) = -\sum_{x\in \mathcal{X}} n_x G^\ast_x\left(%
\frac{\bm{\mu}_{x\cdot}}{n_x}\right) -\sum_{y\in \mathcal{Y}} m_y
H^\ast_y\left(\frac{\bm{\mu}_{\cdot y}}{m_y}\right).
\end{equation}
The function $\mathcal{E}$ only depends on the matching patterns $\bm{\mu}$
and the margins $\bm{q}= (\bm{n},\bm{m})$. It is concave; its shape depends
on the distributions $(\mathbb{P}_{x})$ and $(\mathbb{Q}_{y})$ of the unobserved
heterogeneity terms $\bm{\varepsilon}$ and $\bm{\eta}$.

\subsection{The Data}

We assume that the analyst observes a random sample of size $N$ from a large
population of households. By simple counting (possibly using sampling
weights), she obtains estimators of the matching patterns $\hat{\mu}_{xy}, 
\hat{\mu}_{x0}$, and $\hat{\mu}_{0y}$, as well as the margins: 
\begin{align*}
\hat{n}_x &= \hat{\mu}_{x0}+\sum_{y\in \mathcal{Y}}\hat{\mu}_{xy} \\
\hat{m}_y &= \hat{\mu}_{0y}+\sum_{x\in \mathcal{X}}\hat{\mu}_{xy}
\end{align*}
and a consistent estimator $\Sigma_{\hat{\bm{\mu}}}$ of their asymptotic variance-covariance
matrix, given by
$$ \Sigma_{\hat{\bm{\mu}}} = \text{diag}(\hat{\bm{\mu}} )-\hat{\bm{\mu}} \hat{\bm{\mu}}^\top. $$

\section{Minimum-distance Estimation}

Recall that we have assumed that each $\mathbb{P}_x$ (resp.\ each $\mathbb{Q}%
_y$) has full support on $\mathrm{I\kern-.17em R}^{Y+1}$ (resp.\ $\mathrm{I%
\kern-.17em R}^{X+1}$). Then all $\mu_{xy}, \mu_{x0}, \mu_{0y}$ must be
positive; as a consequence, the $G_x, H_y, G^\ast_x, H^\ast_y$ functions are
continuously differentiable everywhere, as is the generalized entropy
function $\mathcal{E}$.

\noindent \citecomplete{cupid:restud} showed that at the stable matching $\bm{\mu}$, the
joint surplus matrix $\bm{\Phi}$ can be obtained by the following simple
formula: 
\begin{equation}
\Phi _{xy}=-\frac{\partial \mathcal{E}}{\partial \mu _{xy}}(\bm{\mu},\bm{q}).
\label{IdentPhi}
\end{equation}
These are the first-order conditions of the maximization of the total joint
surplus  
\begin{equation*}
\mathcal{W} = \max_{\bm{\mu}} \left(\sum_{x,y} \mu_{xy}\Phi_{xy}+ \mathcal{E}%
(\bm{\mu},\bm{q})\right).  
\end{equation*}

Suppose that the distributions $\mathbb{P}_x$ and $\mathbb{Q}_y$ are
specified up to a parameter vector $\bm{\alpha} \in \mathbb{R}^{d_{\bm{\alpha}}}$,
while the joint surplus matrix $\bm{\Phi}$ is specified up to a parameter
vector $\bm{\beta} \in \mathbb{R}^{d_{\bm{\beta}}}$. We write the generalized
entropy function $\mathcal{E}^{\bm{\alpha}}$ and the parameterized surplus vector $\bm \Phi^{\bm \beta}$. Then one can use~%
\eqref{IdentPhi} as the basis for a minimum distance estimator\footnote{Note that in general, one should choose $d_{\bm{\alpha}}+d_{\bm{\beta}} \leq X \times Y$ to ensure identification.}. That is, we
write a mixed hypothesis as 
\begin{equation*}
\exists \bm{\lambda}=(\bm{\alpha}, \bm{\beta}), \; \; \bm{D}^{\bm{\lambda}}(%
\bm{\mu},\bm{q}) \equiv \bm{\Phi}^{\bm{\beta}}+\frac{\partial \mathcal{E}^{%
\bm{\alpha}}}{\partial \bm{\mu}}(\bm{\mu}, \bm{q})=\bm{0},
\end{equation*}
stacking all $X\times Y$ conditions in~\eqref{IdentPhi} in a vector $\bm{D}^{%
\bm{\lambda}}$.

We choose $\hat{\bm{\lambda}}$ to minimize $\norm{\bm{D}^{\bm{\lambda}}(\bm{%
\hat{\mu}}, \bm{\hat{q}})}^2_{\bm{S}}$ for some positive definite $(X\times
Y, X\times Y)$ matrix $\bm{S}$. By the general theory of minimum distance
estimators, we know that this yields a consistent estimator of $\bm{\lambda}$
if the model is well specified, and that if we choose $\bm{S}=%
\bm{\hat{\Omega}}^{-1}$ where $\bm{\hat{\Omega}}$ consistently estimates $V%
\bm{D}^{\bm{\lambda}}(\bm{\hat{\mu}},\bm{\hat{q}})$ (and can be obtained by the delta method), the minimum distance
estimator will reach its efficiency bound. Further, if the model is well
specified and the choice of $\bm{S}$ is the efficient one, the minimized
value of the squared norm follows a $\chi^2$ of degree $X\times
Y-d_{\bm{\alpha}}-d_{\bm{\beta}}$. Note that this optimization problem is \emph{not} a convex optimization problem in general.

\subsection{The Linear Case}

\label{sub:mde:linear}

Minimum-distance estimation is a particularly appealing strategy if both the
derivatives of the generalized entropy function $\mathcal{E}^{\bm{\alpha}}$ and the
surplus matrix $\bm{\Phi}^{\bm{\beta}}$ are linear in the parameters: 
\begin{equation}  \label{eq:dElinear}
\frac{\partial \mathcal{E}^{\bm{\alpha}}}{\partial \mu_{xy}}(\bm{\mu}, \bm{q}%
) = e^0_{xy}(\bm{\mu}, \bm{q})+ e_{xy}(\bm{\mu}, \bm{q}) \cdot \bm{\alpha}
\end{equation}
and 
\begin{equation}  \label{eq:Philinear}
\Phi_{xy}^{\bm{\beta}}=\bm{\phi}_{xy}\cdot \bm{\beta}
\end{equation}
for some vectors of basis functions $\bm{e}(\bm{\mu}, \bm{q})$ and $\bm{\phi}
$. Then 
\begin{equation*}
D_{xy}^{\bm{\lambda}}(\bm{\mu}, \bm{q})=\bm{\phi}_{xy}\cdot \bm{\beta}+
e^0_{xy}(\bm{\mu}, \bm{q})+e_{xy}(\bm{\mu}, \bm{q}) \cdot \bm{\alpha} 
\end{equation*}
is linear in the parameters $\bm{\lambda}$. (Recall that, for every $(x,y) \in \mathcal X \times \mathcal Y$, the vector $ e_{xy}(\bm{\mu}, \bm{q})$ is of size $d_{\bm{\alpha}}$, and $\bm {\phi}_{xy}$ is of size $d_{\bm \beta}$.)

These two conditions call for several remarks. Condition~\eqref{eq:Philinear}
is a natural choice for a flexible specification. Condition~%
\eqref{eq:dElinear} trivially holds in models where the $\mathbb{P}_x$ and $%
\mathbb{Q}_y$ are parameter-free, like the ubiquitous~\citecomplete{choo-siow:06}
specification. As we will see, it holds in several other leading examples.
Note also that the parameter-free part $\bm{e}^0$ is necessary in order to
normalize the scale of the error terms, which is otherwise not identified in
this discrete-choice model.

Under conditions~\eqref{eq:dElinear} and~\eqref{eq:Philinear}, the minimum
distance estimator can be implemented by linear least-squares. Let $%
\bm{\hat{F}}$ denote the $(X\times Y, d_{\bm{\alpha}}+d_{\bm{\beta}})$ matrix that stacks $%
\bm{e}(\bm{\hat{\mu}},\bm{\hat{q}})$ and $\bm{\phi}$ vertically, so that $%
\bm{D}^{\lambda}(\bm{\hat{\mu}},\bm{\hat{r}})=\bm{\hat{e}}^0+\bm{\hat{F}}%
\bm{\lambda}$, where $\bm{\hat{e}}^0 = \bm{e}^0(\bm {\hat{\mu}},\bm{\hat{q}} ) $ Then for any choice of $\bm{S}$, the minimum distance
estimator $\bm{\hat{\lambda}}$ solves the linear system 
\begin{equation}  \label{eq:solvelambda}
\left(\bm{\hat{F}}^\top \bm{S}\bm{\hat{F}}\right) \; \hat{ \bm{\lambda} } = -%
\bm{\hat{F}}^\top \bm{S}\bm{\hat{e}}^0.
\end{equation}
Since $\bm{\hat{e}}^0$ and $\bm{\hat{F}}$ are functions of $(\bm{\hat{\mu}}, %
\bm{\hat{q}})$, the variance $\bm{\hat{\Omega}}(\bm{\lambda})$ of $%
\bm{\hat{D}}^{\bm{\lambda}}$ can be computed from $\bm{\hat{V}}$ using the
delta method. Again, taking $\bm{S}$ to be the inverse of $\bm{\hat{\Omega}}(%
\bm{\hat{\lambda}})$ is the efficient choice. This procedure is summarized
in Box~\hyperref[min-distance-estimation-full]{1}.

\begin{center}
\label{min-distance-estimation-full} 
\begin{tikzpicture}\label{tikz:box1}

\node [interpretationBox] (box){ \begin{minipage}{\textwidth}

\begin{enumerate}
\item Choose any positive definite matrix $\bm{S}$ and  solve~\eqref{eq:solvelambda} for a consistent estimator $\bm{\lambda}^\ast$
\item Use the delta method to estimate the  variance $\bm{\Omega}^\ast$ of $\bm{\hat{D}}^{\bm{\lambda}}$ at $\bm{\lambda}=\bm{\lambda^\ast}$; let $\bm{S}^\ast=(\bm{\Omega}^\ast)^{-1}$
\item Take $\bm{S}=\bm{S}^\ast$ and solve~\eqref{eq:solvelambda} again for $\bm{\hat{\lambda}}$
\item The  variance-covariance matrix of $\bm{\hat{\lambda}}$ is consistently estimated by
\[
\left(\bm{\hat{F}}^\top \bm{S}^\ast \bm{\hat{F}}\right)^{-1}
\]
\item Under the null of correct specification, the statistic
\[
\bm{\hat{T}}=\left(\bm{\hat{D}}^{\bm{\hat{\lambda}}}\right)^\top \bm{S}^\ast \bm{\hat{D}}^{\bm{\hat{\lambda}}}
\]
converges to a $\chi^2(X \times Y - d_{\bm{\alpha}}-d_{\bm{\beta}})$ distribution.
\end{enumerate}

    \end{minipage}
};
\node[interpretationBoxTitle, right=10pt] at (box.north west) 
{Box 1: min-distance estimation, linear case};

\end{tikzpicture}
\end{center}

If the distributions $(\mathbb{P}_x)$ and $(\mathbb{Q}_y)$ are
parameter-free, the matrix $\bm{\Omega}^\ast$ does not depend on $%
\bm{\lambda}$ any more, and $\bm{\hat{F}}$ is simply the matrix $\bm{\phi}$.
The estimators of $\bm{\lambda}=\bm{\beta}$ can be obtained following the
procedure described in Box~\hyperref[min-distance-estimator]{2}.

\begin{center}
\begin{tikzpicture}\label{tikz:box2}
\label{min-distance-estimator}
\node [interpretationBox] (box){ \begin{minipage}{\textwidth}

\begin{enumerate}
    \item Evaluate $\bm{\Omega}^\ast = V \bm{\hat{e}}^0$ and $\bm{S}^\ast=(\bm{\Omega}^\ast)^{-1}$
    \item Solve the linear system $\left(\bm{\phi}^\top \bm{S}^\ast \bm{\phi}\right) \; \bm{\beta} = -\bm{\phi}^\top \bm{S}^\ast \bm{\hat{e}}^0$
\item The  variance-covariance matrix of $\bm{\hat{\beta}}$ is consistently estimated by
\[
\left(\bm{\phi}^\top \bm{S}^\ast \bm{\phi}\right)^{-1}
\]
\item Under the null of correct specification, the statistic
\[
\bm{\hat{T}}=\left(\bm{\phi}\bm{\hat{\beta}}+\bm{\hat{e}}^0\right)^\top \bm{S}^\ast \left(\bm{\phi}\bm{\hat{\beta}}+\bm{\hat{e}}^0\right)
\]
converges to a $\chi^2(X\times Y - d_{\bm{\beta}})$ distribution.
\end{enumerate}

    \end{minipage}
};
\node[interpretationBoxTitle, right=10pt] at (box.north west) 
{Box 2: min-distance estimator, linear case with parameter-free heterogeneity};

\end{tikzpicture}
\end{center}

Note that since $\bm{\phi}$ is non-random, the variance of $\bm{\hat{D}}^{%
\bm{\lambda}}$ is the variance of the derivative of the generalized entropy.
Step~2 therefore requires evaluating the second derivatives of the
generalized entropy $\mathcal{E}$: by the delta method, 
\begin{equation*}
V\bm{\hat{D}}^{\bm{\lambda}} = 
\begin{pmatrix}
\mathcal{E}_{\bm{\mu}\bm{\mu}}^\top & \mathcal{E}_{\bm{\mu}\bm{q}}^\top%
\end{pmatrix}
V 
\begin{pmatrix}
\bm{\hat{\mu}} \\ 
\bm{\hat{q}}%
\end{pmatrix}
\begin{pmatrix}
\mathcal{E}_{\bm{\mu}\bm{\mu}} \\ 
\mathcal{E}_{\bm{\mu}\bm{q}}%
\end{pmatrix}%
. 
\end{equation*}
It is easy to see from the definition in~\eqref{eq:defE} that the first
derivative of $\mathcal{E}$ with respect to $\mu_{xy}$ only depends on the
conditional matching patterns $\mu_{\cdot\vert x}=(\mu_{x1}/n_x, \ldots,
\mu_{xY}/n_x)$ of men of type $x$, and on those of women of type $y$. As a
consequence, the Hessians of $\mathcal{E}$ are very sparse and are often
easy to evaluate.

\subsection{Examples}

We start with two examples for which the generalized entropy and its
derivatives are available in closed form; in both cases, the derivatives are
linear in the parameters $\bm{\alpha}$. In our third example, the
calculation requires finding the fixed point of a contraction, in a way that
is familiar from empirical industrial organization.

\subsubsection{The Heteroskedastic Logit Model}

Let us start with an easy extension of the~\citecomplete{choo-siow:06} logit model: 
the distributions $\mathbb{P}_x$ and $\mathbb{Q}_y$ are type I-EV iid
vectors with unknown scale factors $\sigma_x$ and $\tau_y$ respectively.
Then $\bm{\alpha}=(\bm{\sigma},\bm{\tau})$ and the derivatives of the
generalized entropy function are linear in $\bm{\alpha}$:  
\begin{equation*}
\frac{\partial \mathcal{E}^{\bm{\alpha}}}{\partial \mu_{xy}}(\bm{\mu}, \bm{q}%
) = -\sigma_x \log\frac{\mu_{xy}}{\mu_{x0}} -\tau_y \log\frac{\mu_{xy}}{%
\mu_{0y}}  
\end{equation*}
where $\mu_{x0}=n_x-\sum_{y\in \mathcal{Y}} \mu_{xy}$ and $%
\mu_{0y}=m_y-\sum_{x\in \mathcal{X}} \mu_{xy}$. The second derivatives of
the generalized entropy take a very simple form:  
\begin{equation}  \label{eq:hessiancshetero:mumu}
\frac{\partial^2 \mathcal{E}^{\bm{\alpha}}}{\partial \mu_{xy}\partial\mu_{zt}%
}(\bm{\mu}, \bm{q}) = -\frac{\sigma_x}{\mu_{x0}}\; \mathrm{1\kern-.40em 1}%
(z=x) -\frac{\tau_y}{\mu_{0y}}\; \mathrm{1\kern-.40em 1}(t=y) -\frac{%
\sigma_x+\tau_y}{\mu_{xy}}\; \mathrm{1\kern-.40em 1}(z=x, t=y)
\end{equation}
and  
\begin{equation}  \label{eq:hessiancshetero:mur}
\frac{\partial^2 \mathcal{E}^{\bm{\alpha}}}{\partial \mu_{xy}\partial n_z}(%
\bm{\mu}, \bm{q}) = \frac{\sigma_x}{\mu_{x0}}\; \mathrm{1\kern-.40em 1}%
(z=x); \; \; \; \; \; \frac{\partial^2 \mathcal{E}^{\bm{\alpha}}}{\partial
\mu_{xy}\partial m_t}(\bm{\mu}, \bm{q}) = \frac{\tau_y}{\mu_{0y}}\; \mathrm{1%
\kern-.40em 1}(t=y).
\end{equation}

Scale normalization is done by fixing the value of one of the parameters in $%
\bm{\alpha}$. The~\citeauthor{choo-siow:06} homoskedastic model obtains when all $%
\sigma_x$ and $\tau_y$ equal one; a gender-heteroskedastic model would have
all $\sigma_x$ equal to one and all $\tau_y$ equal to an unknown $\tau$. 
\citecomplete{csw:17} applied a minimum distance estimator to the homoskedastic and
heteroskedastic logit models.

\subsubsection{Nested Logit}

Consider a two-layer nested logit model. Take men of type $x$ first.
Alternative $0$ (singlehood) is obviously special; we put it alone in its
nest. Each other nest $n\in \mathcal{N}_x$ contains alternatives $y\in 
\mathcal{Y}_n$. The correlation of alternatives within nest $n$ is proxied
by $1-\left(\rho^x_{n}\right)^2$ (with $\rho^x_{0}=1$ for the nest made of
alternative $0$). Similarly, for women of type $y$, alternative $0$ is in a
nest by itself with parameter $\delta^y_0=1$ and alternatives $x\in \mathcal{%
X}_{n^\prime}$ are in a nest $n\in \mathcal{N}^\prime_y$ with parameter $%
\delta^y_{n^\prime}$. We collect the parameters $\bm{\rho}$ and $\bm{\delta}$
into $\bm{\alpha}$.

The formul\ae\ in Example 2.1 of \citecomplete{cupid:restud} imply that if $y$ is in
nest $n \in \mathcal{N}_x$ and $x$ is in nest $n^\prime \in \mathcal{N}_y$,
then 
\begin{align}  \label{eq:nested_logit}
\frac{\partial \mathcal{E}^{\bm{\alpha}}}{\partial \mu_{xy}}(\bm{\mu}, \bm{q}%
) &= -\rho^x_n \log\frac{\mu_{xy}}{\mu_{x0}} - (1-\rho^x_n) \log\frac{%
\mu_{xn}}{\mu_{x0}}  \notag \\
& -\delta^y_{n^\prime} \log\frac{\mu_{xy}}{\mu_{0y}} -
(1-\delta^y_{n^\prime}) \log\frac{\mu_{n^\prime y}}{\mu_{0y}},
\end{align}
where we defined $\mu_{xn}=\sum_{t\in \mathcal{Y}_n} \mu_{xt}$ and $%
\mu_{n^\prime y}=\sum_{z\in \mathcal{X}_{n^\prime}} \mu_{zy}$. Once again,
this is linear in the parameters $\bm{\alpha}$; it remains linear if we
impose constraints on the nests (for instance, that $\mathcal{N}_x$ is the
same for all types $x$) and/or linear constraints on the $\bm{\rho}$
parameters (for instance, that $\rho_{xn}$ only depends on $n$).

\subsubsection{Mixed Logit}

Let us now describe a random coefficient logit model. Consider a man $i$ of
type $x$, endowed with preferences $\bm{e}_i$ over a set of $d$ observable
characteristics $\bm{Z}$ of potential partners. We add an idiosyncratic
shock $\bm{\zeta}_i$ that is distributed as a standard iid type I extreme
value vector over $\mathrm{I\kern-.17em R}^{Y+1}$, independently of $\bm{e}_i
$, and a scale factor $s>0$: 
\begin{equation*}
\varepsilon_{iy}= \sum_{k=1}^d Z_{yk}e_{ik} + s\zeta_{iy} 
\end{equation*}
or in matrix form: $\bm{\varepsilon}=\bm{Z} \bm{e}+s\bm{\zeta}$. This
specification is standard in empirical IO. In \citecomplete{blp:1995}: the covariates
in $\bm{Z}$ stand for the observed characteristics of the products; the $%
\bm{e}$ are individual valuations of these characteristics, and the $\bm{\zeta}$
are idiosyncratic shocks.

Let individual preferences $\bm{e}$ of men of type $x$ have distribution $%
\mathbb{P}^e_{x}$. We will seek to estimate the parameters $\bm{\beta}$ of
the joint surplus, the scale factor $s$, and the parameters of the
distributions $\mathbb{P}^e_{x}$. We collect $s$ and the parameters of $%
\mathbb{P}^e_{x}$ in a vector $\bm{\alpha}$.

To compute the derivative of the generalized entropy function, we recall
from~\citecomplete{cupid:restud} that 
\begin{equation*}
G^{\ast}_x(\bm{\nu};\bm{\alpha})= -\min_{U_0=0, \bm{U} \in \mathbb{R}^Y} \left[ \int s \log\sum_{y=0, 1, \ldots, Y} \exp\left(\frac{%
U_{y}+\left(\bm{Z}\bm{e}\right)_{y}}{s}\right) d\mathbb{P}^e_{x}(\bm{e})
-\sum_{y\in \mathcal{Y}} \nu_{y}U_{y}\right].
\end{equation*}%
By the envelope theorem, the derivative of $G^{\ast}_x(\bm{\nu};\bm{\alpha})$
with respect to $\bm{\nu}$ is the vector $\bm{U}$ that solves the system 
\begin{equation*}
\nu_y = \int \frac{\exp((U_{y}+\bm{Z}_y\bm{e})/s)}{\sum_{t=0, 1, \ldots, Y}
\exp((U_{t}+\bm{Z}_t\bm{e})/s)} d\mathbb{P}^e_{x}(\bm{e}) \; \; \forall y=1,
\ldots, Y. 
\end{equation*}
This is exactly isomorphic to the inversion problem in~\citecomplete{blp:1995}, with
the unknown $\bm{U}$ standing for the product effects and $\bm{\nu}$ playing
the role of the product market shares. After replacing $\bm{\nu}$ with the
observed $\bm{\mu}_{x\cdot}/n_x$, the system can be solved by any of the
algorithms that are standard in this literature. The solution gives row $x$
of the matrix $\bm{U}$. Proceeding in the same way for other types of men,
and solving for $\bm{V}$ for women, gives the derivatives of the generalized
entropy function: 
\begin{equation*}
\frac{\partial \mathcal{E}^{\bm{\alpha}}}{\partial \mu_{xy}}(\bm{\mu}, \bm{q}%
) = - \frac{\partial G^\ast_x}{\partial \nu_{xy}}\left(\frac{\bm{\mu}%
_{x\cdot}}{n_x}\right) - \frac{\partial H^\ast_y}{\partial \nu_{xy}}\left(%
\frac{\bm{\mu}_{\cdot y}}{m_y}\right) = -U_{xy}-V_{xy}. 
\end{equation*}
The limit case $s=0$ yields the pure characteristics model of~%
\citecomplete{berrypakes:07}. Then the system to be solved for row $x$ of $\bm{U}$
is 
\begin{equation*}
\nu_y = \mathbb{P}^e_{x}\left(y \in\arg\max_{t=0, 1, \ldots, Y} (U_{t}+\bm{Z}_t%
\bm{e})\right) \; \; \forall y=1, \ldots, Y. 
\end{equation*}
If each $\bm{Z}_t$ is a scalar, the inequalities boil down to 
\begin{equation*}
e^-(y; \bm{U},\bm{Z})\equiv \max_{t \vert Z_t < Z_y} \frac{U_t-U_y}{Z_y-Z_t}
\leq e \leq \min_{t \vert Z_t > Z_y} \frac{U_y-U_t}{Z_t-Z_y} \equiv e^+(y;%
\bm{U},\bm{Z}), 
\end{equation*}
and the system of equations to be solved for $\bm{U}$ is 
\begin{equation*}
\nu_y = \mathbb{P}^e_{x}(e^+(y;\bm{U},\bm{Z}))-\mathbb{P}^e_{x}(e^-(y;\bm{U},\bm{Z})) \; \; \forall
y=1, \ldots, Y. 
\end{equation*}

\section{Moment-based Estimation by Poisson Regression}

\label{sec:poisson}

Now take the generalized entropy function $\mathcal{E}$ as known/assumed;
and assume that the joint surplus vector $\bm{\Phi} \in \mathbb{R}^ {%
\mathcal{X} \times \mathcal{Y} } $ is semilinear: $\bm{\Phi}^{\bm{\beta}}=%
\bm{\phi} \bm{\beta}$, where $\bm{\beta}$ is a vector of dimension $d_{\bm{\beta}}$ and $%
\phi$ is a $| \mathcal{X}| |\mathcal{Y}| \times d_{\bm{\beta}}$ matrix. \citecomplete{cupid:restud} introduced a moment-matching procedure that gives a consistent
estimator of the parameter vector $\bm{\beta}$ if the model is
well-specified. The \emph{moment matching estimator\/} equalizes the
observed and simulated \emph{comoments}, that is the expectations of the
basis functions $\bm{\phi}$ under the observed and simulated matching
patterns: 
\begin{equation*}
\sum_{x,y} \hat{\mu}_{xy} \bm{\phi}_{xy}= \sum_{x,y} \mu^{\bm{\beta}}_{xy} %
\bm{\phi}_{xy},
\end{equation*}
where $\bm{\mu}^{\bm{\beta}}$ denotes the stable matching patterns for the
parameter vector $\bm{\beta}$. As explained in~\citecomplete{cupid:restud}, these are
the first-order conditions of the following maximization problem: 
\begin{equation}  \label{eq:mm}
\max_{\bm{\beta}} \left( \bm{\hat{\mu}} \bm{\Phi}^{\bm{\beta}} -\mathcal{W}( %
\bm{\beta},\bm{q} ) \right)
\end{equation}
where $\mathcal{W}(\bm{\beta},\bm{q})=\max_{\bm{\mu}}\left(\bm{\mu} \bm{\Phi}%
^{\bm{\beta}} +\mathcal{E}(\bm{\mu},\bm{q})\right)$ is the value of the
total joint surplus. With a semilinear specification for $\bm{\Phi}^{%
\bm{\beta}}$, both of these problems are globally convex.

We now show that in the specific (but popular) case of the~\citecomplete{choo-siow:06} model, moment matching can be reformulated as a generalized
linear model, and estimated by a Poisson regression with two-sided fixed
effects.

Define $\mathcal{A=X\times Y\cup X\times }\left\{ 0\right\} \cup \left\{
0\right\} \times \mathcal{Y}$ the set of possible marital arrangements.
Define a vector $w \in \mathbb{R}^\mathcal{A}$ by $w_{xy}=2$ if $x\in 
\mathcal{X} \text{ and } y\in \mathcal{Y}$ and $w_{xy} = 1$ if $x = 0$ or if 
$y=0$, so that $w_{xy}$ is the size of household $xy$, namely 2 if matched,
1 if single. The following theorem summarizes our results.

\begin{theorem}[Estimating the logit model with a Poisson regression]
\label{thm:mmlogit} In the \citeauthor{choo-siow:06} model, the moment-matching estimator $%
\hat{\beta}$ is the solution to a Poisson regression of $\left( \hat{\mu}%
_{xy}\right) _{xy\in \mathcal{A}}$ on $\left( \Phi _{xy}^{\beta
}/w_{xy}\right) _{xy\in \mathcal{A}}$, with with $x$- and $y$- fixed effects
and with weights $w_{xy}$ defined above, and where we take by convention $%
\Phi _{x0}^{\beta }=0$ and $\Phi _{0y}^{\beta }=0$ and $a_{0}=0$ and $b_{0}=0
$. In other words, $\beta $ is the solution to%
\begin{equation*}
\max_{\beta _{k},a_{x},b_{y}}\sum_{xy\in \mathcal{A}}w_{xy}\hat{\mu}%
_{xy}\left( \frac{\Phi _{xy}^{\beta }-a_{x}-b_{y}}{w_{xy}}\right)
-\sum_{xy\in \mathcal{A}}w_{xy}\exp \left( \frac{\Phi _{xy}^{\beta
}-a_{x}-b_{y}}{w_{xy}}\right) . 
\end{equation*}
\end{theorem}

\bigskip

The proof of Theorem~\ref{thm:mmlogit} is given in Appendix~\ref%
{appx:proof_thm}. The result is very useful in that it allows for inference
on $\bm{\beta},\bm{u}$ and $\bm{v}$ in semilinear logit models with standard
statistical packages such as~\texttt{glm} in R, or~\texttt{scikit-learn} in
Python. Note that like \citecomplete{sstgravity:06} in the international trade
literature, we end up fitting a Poisson regression to a model that is
definitely not generated by a Poisson count process. The motivation is
different, however. They start from a semiparametric model of the gravity
equation and use the robustness of the Poisson pseudo-maximum likelihood
estimator. We start from a more complex, fully specified structural model
and we show that a semiparametric estimator (moment-matching) is numerically
equivalent to the maximum likelihood estimator of a Poisson model.

In the sequel we will denote $\bm{I}_m$ the $(m,m)$ identity matrix; $\bm{p}%
_{(m,n)}$ the $(m,n)$ matrix whose elements all equal $p$; and $\bm{p}%
_m\equiv \bm{p}_{(m,1)}$. Also, we say that we stack an $(X,Y)$ matrix in
``row-major order'' when we create a vector of $X\times Y$ elements whose
first $Y$ elements are the first row of the matrix, etc.

\begin{center}
\begin{tikzpicture}\label{tikz:box3}
\label{GLM-estimator}
\node [interpretationBox] (box){ \begin{minipage}{\textwidth}

\begin{enumerate}
\item Flatten the observed matching patterns $\bm{\hat{\mu}}$ into a vector of size $|\mathcal{A}|$, by first stacking the elements $xy \in \mathcal{X} \times \mathcal{Y}$ in row-major order, then adding the elements $x0 \in \mathcal{X}\times \{0\}$, and finally adding the elements $0y \in  \{0\} \times \mathcal{Y}$. 
\item For each basis function $k=1,\ldots ,K$%
, represent the vector$(\bm{\phi}_{xy}^{k})_{xy \in \mathcal{A}}$ in the same order. Then
represent the $|\mathcal{A}| \times K$ matrix $\phi$ from these $K$ column vectors of size $ |\mathcal{A}|$.
\item Using the same order again, represent the vector $\bm{w}$ in $\mathbb{R}^\mathcal{A}$:
\begin{equation*}
\bm{w}=(\bm{2}_{X\times Y}^\top, \bm{1}_X^\top, \bm{1}_Y^\top)^\top.
\end{equation*}%
\item Finally, define the $|\mathcal{A}| \times (X+Y+K)$ matrix $\bm{Z}$ as
\begin{equation*}
\bm{Z}=\begin{pmatrix}
\bm{\phi}/2 & -\frac{1}{2}\bm{I}_{X}\otimes \bm{1}_{(Y,1)} & -%
\frac{1}{2}\bm{1}_{(X,1) }\otimes \bm{I}_{Y}   \\ \bm{0}_{(X,K)} &
 -\bm{I}_{X} & \bm{0}_{(X,Y)}  \\  \bm{0}_{(Y,K)} &
\bm{0}_{(Y,X)} & -\bm{I}_{Y}  &
\end{pmatrix}.
\end{equation*}%

\item Run a Poisson regression of $\hat{\mu}$ on $\bm{Z}$ with weights $\bm{w}$. Do not add fixed effect, as these have already been included in the design of $\bm{Z}$. Let $\bm{\hat{\gamma}}$ be the vector of coefficients obtained this way; it solves
\[
\max_{\gamma \in \mathbb{R}^{K+X+Y}}\left(\sum_{a\in \mathcal{A}}w_{a}\hat{\mu}%
_{a}\left( Z\gamma \right) _{a}-\sum_{a\in \mathcal{A}}w_{a}\exp \left(
\left( Z\gamma \right) _{a}\right)\right).
\]

\item Decompose $\bm{\hat{\gamma}}=(\bm{\hat{\beta}}^\top, \bm{\hat{a}}^\top,\bm{\hat{b}}^\top)^\top \in \mathbb{R}^{K+X+Y}$. Then $\bm{\hat{\beta}}$ is the moment-matching estimator, and $a_x$ and $b_y$ are the $x$- and $y$- fixed effects.

\end{enumerate}

    \end{minipage}
};
\node[interpretationBoxTitle, right=10pt] at (box.north west) 
{Box 3: GLM estimator, linear case with logit heterogeneity};

\end{tikzpicture}
\end{center}

As a result, we get that:

\begin{theorem}\label{thm:mmlogit-var} The asymptotic variance-covariance matrix of $\bm{\hat{\gamma}}$
can be estimated with 
\begin{equation*}
\hat{V}\hat{\bm{\gamma}}=\bm{\hat{A}}^{-1}\;\hat{\bm{B}}\;\bm{\hat{A}}^{-1}
\end{equation*}%
where, letting $\bm{W}=\text{diag}\left( w\right) $, we have 
\begin{eqnarray*}
\bm{\hat{A}} &=&\left( \bm{Z}^{\top }\bm{W}\text{diag}\left( \exp \left( \bm{Z}%
\gamma \right) \right) \bm{Z}\right)  \\
&=&\sum_{a\in \mathcal{A}}w_{a}\exp (\bm{Z}_{a}\hat{\bm{\gamma}})\bm{Z}%
_{a}^{\top }\bm{Z}_{a}
\end{eqnarray*}%
and 
\begin{eqnarray*}
\bm{\hat{B}} &=&\bm{Z}^{\top }\bm{W}(\text{diag}(\hat{\mu})-\hat{\mu}\hat{\mu}%
^{\top })\bm{W}\bm{Z} \\
&=&\sum_{a\in \mathcal{A}}w_{a}\hat{\mu}_{a}\bm{Z}_{a}^{\top }\bm{Z}%
_{a}-\sum_{a,a^{\prime }\in \mathcal{A}}w_{a}w_{a^{\prime }}\;\hat{\mu}_{a}%
\hat{\mu}_{a^{\prime }}\;\bm{Z}_{a}^{\top }\bm{Z}_{a^{\prime }}.
\end{eqnarray*}
\end{theorem}

\section{Monte Carlo Simulation}

\label{sec:monte_carlo} We coded these two estimation methods in a \texttt{%
Python} package called \texttt{cupid\_matching} that is available from the
standard repositories\footnote{%
See \url{https://pypi.org/project/cupid-matching/}.}. To test the quality of
the estimators, we generated data both from a  \citeauthor{choo-siow:06}
model and from a semilinear nested logit model. We use both the Poisson
estimator and the minimum-distance estimator on the former model, and only
the minimum-distance estimator of course on the latter.

In both cases, we take $X=Y=20$ and we use $K=8$ basis functions: $1, x, y,
x^2, xy, y^2, \mathrm{1\kern-.40em 1}(x\geq y), \max(x-y, 0)$. The true
data-generating process has 
\begin{equation*}
\Phi_{xy} = 1-\frac{(x-y)^2}{100} + 0.5\mathrm{1\kern-.40em 1}(x\geq y), 
\end{equation*}
so that the true $\bm{\beta}$ is $(1.0,0.0,0.0,-0.01,0.02,-0.01,0.5, 0.0)$.
This could be interpreted as the joint surplus from marriage as a function
of the ages of the husband $x$ and of the wife $y$. It is highest when the
partners have the same age; if they don't, it is larger when the husband is
the older partner. We use equal numbers of men and women; and we choose
vectors $\bm{n}=\bm{m}$ whose elements form a decreasing geometric sequence
with rate $0.8$ (there are fewer individuals available for marriages at
higher ages).

\subsection{Semilinear Logit}

The semilinear logit model is entirely described
above. We use the IPFP algorithm described in Section~4.2 of \citecomplete{cupid:restud}
to solve for the stable matching patterns $\bm{\mu}$ for the margins $\bm{n}$
and $\bm{m}$. To generate a sample, we draw randomly $N=10,000$ households
from the multinomial probability distribution generated by $\bm{\mu}$. We
generated $S=1,000$ such samples. We used minimum distance estimation and
Poisson GLM on each sample. While the minimum distance estimator only uses a
linear regression, the Poisson GLM method uses numerical optimization under
the hood. In our simulations using the \texttt{sklearn} Python package, the
algorithm went astray on 50 of our 1,000 samples, mostly because of overflow
errors. We discarded these samples from our analysis.

\begin{figure}[tbp]
\centering
\includegraphics[height=0.9\textheight]{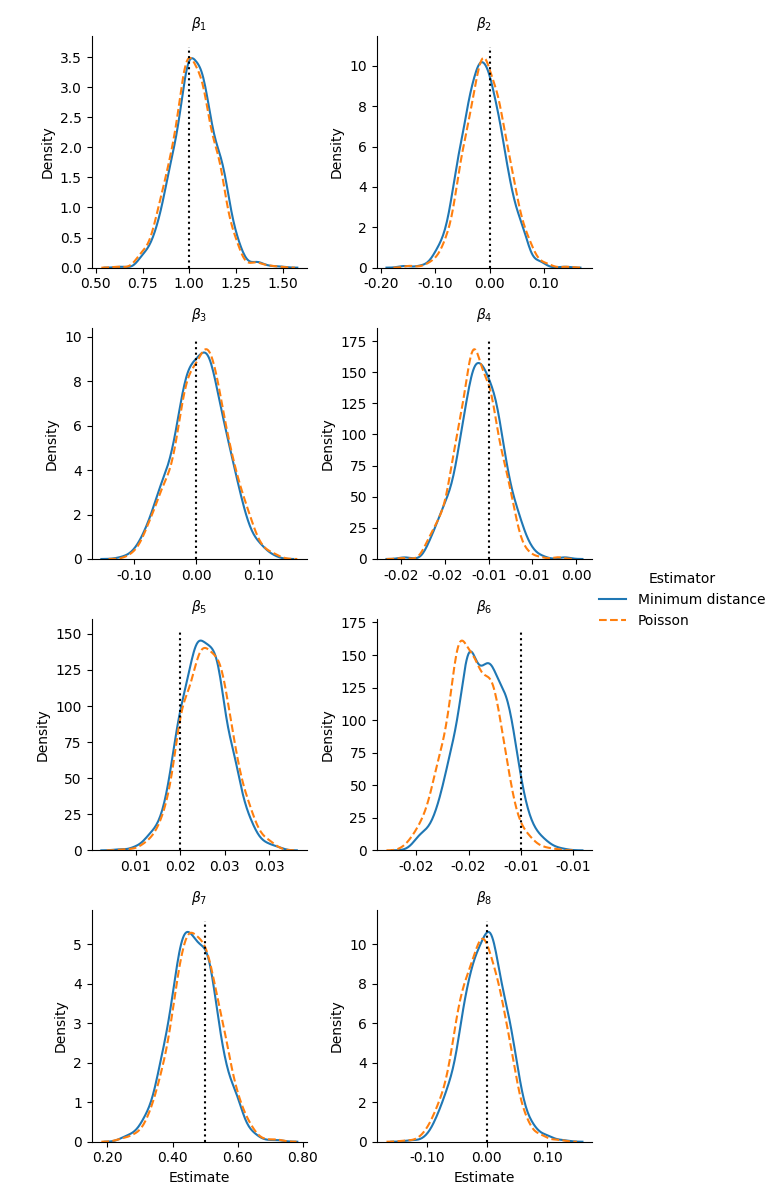}  
\caption{Estimating the Choo and Siow Model}
\label{fig:choo_siow_simul_results_N10000}
\end{figure}

As Figure~\ref{fig:choo_siow_simul_results_N10000} shows, on the remaining
950 samples the two estimators perform about equally well. Both estimators
exploit the same $X\times Y$ moment conditions 
\begin{equation*}
\bm{\phi}\cdot \bm{\beta} + \frac{\partial \mathcal{E}}{\partial \bm \mu} = %
\bm{0}, 
\end{equation*}
and both minimize a quadratic form of these conditions. The difference is in
the weighting matrix. We saw in Section~\ref{sub:mde:linear} that the
minimum-distance estimator uses the variance of the derivative of the
entropy at the observed matching. On the other hand, the Poisson estimator
uses the Hessian of the entropy at the current parameter values. While the
two estimators are quite close in our simulations, one can imagine
situations in which the divergence would be larger.

\subsection{Semilinear Nested Logit}

\label{sub:mc_nested} For both men and women, we defined three nests that
consist of $\{0\}$, $\{1,\ldots,10\}$, and $\{11,\ldots,20\}$. We take the
true nest parameters to be all equal to $0.5$ (that is, $\rho_n^x=\delta_{n^%
\prime}^y=0.5$ for all $n,n^\prime: x\in n$ and $y\in n^\prime$).

To generate samples from the  nested logit model, we proceed as with the
logit model. The only difference is that setting up the system to be solved
for equilibrium requires a bit more work. We describe our IPFP algorithm in
Appendix~\ref{appx:IPFP_nested_logit}.

The minimum distance estimator converges fast on all samples. However, we
found that a sample size of 10,000 households was much too small to get
reliable estimates of the parameters.  Figure~\ref%
{fig:nested_logit_results}
gives the distribution of the estimates of the four nest parameters (first
two rows) and the eight coefficients of the bases for larger sample sizes:
respectively $N=100,000$ and $N=1,000,000$. There is a clear downwards bias
on the nest parameters $\rho$ and $\delta$ when $N=100,000$, to the point that
some estimates are negative. Some of the coefficients of the bases are also
badly estimated. With $N=1,000,000$, the minimum distance estimator performs
much better.

\begin{figure}[tbp]
\centering
\includegraphics[height=0.9\textheight]{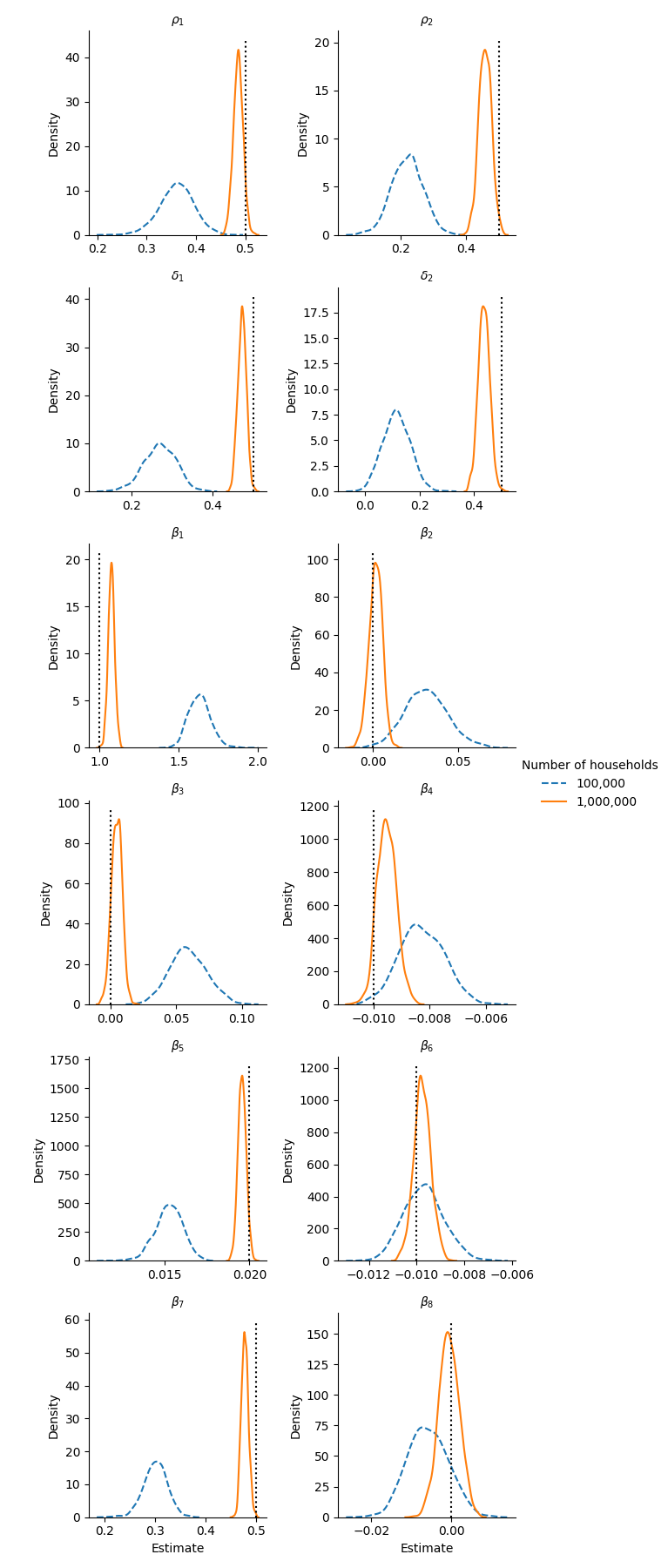}  
\caption{Estimating the Nested Logit Model}
\label{fig:nested_logit_results}
\end{figure}

\section*{Concluding remarks}

Each of the two methods we presented here has its pros and cons.

The minimum-distance estimator applies to all separable models; it is most
convenient in semilinear models. To achieve maximum efficiency, and to test
the specification, one needs to evaluate the second derivatives of the
entropy with respect to the matching patterns. This may be difficult. In
addition, the data often contains zero cells---some $\hat{\mu}_{xy}$ may be
zero. Then the corresponding equation in~\eqref{IdentPhi} is only an
inequality and it must be dropped from the system of estimating equations.
An alternative is to add a small positive number $\delta$ to each $\hat{\mu}%
_{xy}$, to increase the margins $\hat{n}_x$ and $\hat{m}_y$ accordingly, and
to estimate on this adjusted data.

The Poisson regression estimator only applies to semilinear \citecomplete{choo-siow:06} models. It is appealing in its simplicity of use, as one can
rely on standard statistical packages. It is also more robust to zero cells:
nothing in Section~\ref{sec:poisson} relied on taking derivatives with
respect to $\bm{\mu}$ at the observed matching patterns.

Our simulations suggest that it takes large sample sizes to get reliable
estimates of distributional parameters (our $\bm{\alpha}$). In labor markets
or in marriage markets, large samples are readily available. When they are
not (as with matching between firms), it may be better to stick to the \citecomplete{choo-siow:06} specification. Fortunately, the simulations reported in \citecomplete%
{cns:19} are encouraging as to its robustness.

\newpage
\printbibliography
\newpage

\begin{appendix}

\section{IPFP for the Nested Logit}\label{appx:IPFP_nested_logit}
Let us consider a nested logit model in which the nests do not depend on the type ($\mathcal{N}_x\equiv \mathcal{N}$ and $\mathcal{N}^{\prime}_y\equiv \mathcal{N}^\prime$) 
and their parameters $\bm{\rho}$ and $\bm{\delta}$ only depend on  the nest: $\rho^x_n\equiv\rho_n$ and $\delta^y_{n^\prime}\equiv \delta_{n^\prime}$. Equation~\eqref{eq:nested_logit} can be rewritten as follows, for $y\in n$ and $x\in n^\prime$:
\begin{equation}\label{eq:nested_muxy}
\mu_{xy}^{\rho_n+\delta_{n^\prime}} = \exp(\Phi_{xy})\mu_{x0}\mu_{0y} \mu_{xn}^{\rho_n-1}\mu_{n^\prime y}^{\delta_{n^\prime}-1}.
\end{equation}
Since $\mu_{xn}=\sum_{y\in n} \mu_{xy}$, we get
\[
\mu_{xn} = \mu_{x0}^{1/(\rho_n+\delta_{n^\prime})}\mu_{xn}^{(\rho_n-1)/(\rho_n+\delta_{n^\prime})}\sum_{y\in n}
\exp\left(\Phi_{xy}/(\rho_n+\delta_{n^\prime})\right)\mu_{0y}^{1/(\rho_n+\delta_{n^\prime})}
\mu_{n^\prime y}^{(\delta_{n^\prime}-1)/(\rho_n+\delta_{n^\prime})},
\]
and, denoting $K_{xy}=\exp\left(\Phi_{xy}/(\rho_n+\delta_{n^\prime})\right)$:
\begin{equation}\label{eq:nested_muxn}
\mu_{xn}^{(\delta_{n^\prime}+1)/(\rho_n+\delta_{n^\prime})} = \mu_{x0}^{1/(\rho_n+\delta_{n^\prime})}\sum_{y\in n} K_{xy} \mu_{0y}^{1/(\rho_n+\delta_{n^\prime})} \mu_{n^\prime y}^{(\delta_{n^\prime}-1)/(\rho_n+\delta_{n^\prime})}.
\end{equation}
Substituting in the adding up constraint $\mu_{x0}+\sum_{y=1}^Y \mu_{xy}=n_x$ gives
\begin{align}\label{eq:nested_addingup}
n_x &= \mu_{x0} +\sum_{n\in\mathcal{N}} \mu_{xn} \nonumber\\
&= \mu_{x0}+\sum_{n\in \mathcal{N}} 
\mu_{x0}^{1/(\delta_{n^\prime}+1)} 
\left(
\sum_{y\in n} K_{xy} \mu_{0y}^{1/(\rho_n+\delta_{n^\prime})} \mu_{n^\prime y}^{(\delta_{n^\prime}-1)/(\rho_n+\delta_{n^\prime})}
\right)^{(\rho_n+\delta_{n^\prime})/(\delta_{n^\prime}+1)}.
\end{align}
For given values of  $(\mu_{0y},\mu_{n^\prime y})$ for all $y$, \eqref{eq:nested_addingup} defines  $\mu_{x0}$ uniquely\footnote{Since $\delta_{n^\prime}\geq 0$, the right-hand side is an increasing function of $\mu_{x0}$ whose values go from zero to infinity.} . Once $\mu_{x0}$ is known, we can plug it in~\eqref{eq:nested_muxn} to obtain the values of $\mu_{xn}$ for all $n$. We do this for all values of $x$.

Then we can apply similar equations to the $y$ side:
\begin{align*}
    \mu_{n^\prime y}^{(\rho_n+1)/(\rho_n+\delta_{n^\prime})} &= \mu_{0y}^{1/(\rho_n+\delta_{n^\prime})}\sum_{x\in n^{\prime}} K_{xy} \mu_{x0}^{1/(\rho_n+\delta_{n^\prime})} \mu_{xn}^{(\rho_{n}-1)/(\rho_n+\delta_{n^\prime})}\\
    m_y &= \mu_{0y}+\sum_{n^\prime\in \mathcal{N}^\prime} 
\mu_{0y}^{1/(\rho_n+1)} 
\left(
\sum_{x\in n^\prime} K_{xy} \mu_{x0}^{1/(\rho_n+\delta_{n^\prime})} \mu_{xn}^{(\rho_n-1)/(\rho_n+\delta_{n^\prime})}
\right)^{(\rho_n+\delta_{n^\prime})/(\rho_n+1)}
\end{align*}
to solve for $\mu_{0y}$ and $\mu_{n^\prime y}$ given the values of $(\mu_{x0},\mu_{xn})$ for all $x$. We iterate until convergence and we use~\eqref{eq:nested_muxy} to compute the matching patterns $\mu_{xy}$.

\section{Proofs}\label{appx:proof_thm}

\subsection{Proof of theorem~\ref{thm:mmlogit}}
Recall that
\[
N = \sum_{x,y} \mu_{xy}^{\bm{\beta}}+\sum_{x} \mu_{x0}^{\bm{\beta}}+\sum_{y} \mu_{0y}^{\bm{\beta}}
\]
is the total mass of households in the sample.  For the \citecomplete{choo-siow:06} specification we have at the stable matching $(\bm{\mu},\bm{u},\bm{v})$ for a joint surplus $\bm{\Phi}^{\bm{\beta}}$:
\begin{align}\label{eq:mufromuv}
    \mu_{x0} &= \hat{n}_x \exp(-u_x) \nonumber \\
   \mu_{0y} &= \hat{m}_y \exp(-v_y)\\
    \mu_{xy} &= \sqrt{\hat{n}_x\hat{m}_y} \exp((\Phi_{xy}-u_x-v_y)/2) \nonumber.
\end{align}
Consider the maximization of the following expression:
\begin{align*}
&\sum_{x,y}\hat{\mu}_{xy}\bm{\phi}_{xy} \bm{\beta} 
- 2\sum_{x,y} \sqrt{\hat{n}_x\hat{m}_y}
\exp((\bm{\phi}_{x}\bm{\beta}
-u_x-v_y)/2)\\
&\ -\sum_{x} \hat{n}_x \exp(-u_x)
-\sum_{y} \hat{m}_y \exp(-v_y)
- \sum_x \hat{n}_x u_x-\sum_y \hat{m}_y v_y
\end{align*}
over $\bm u$, $\bm v$, and $\bm \beta$. We see that the first order conditions yield that $\bm \mu$ defined in~\ref{eq:mufromuv} satisfies the margin equations 
\begin{align}
\sum_{y} \mu_{xy} + \mu_{x0} = n_x \\
\sum_{x} \mu_{xy} + \mu_{0y} = m_y
\end{align}
for the first order conditions with respect to $u_x$ and $v_y$, and
\[
\sum_{xy} \mu_{xy} \phi_{xy}^k = \sum_{xy} \hat{\mu}_{xy} \phi_{xy}^k 
\]
for the first order conditions with respect to $\beta_k$.

Now remember that the log-likelihood function of a Poisson count model with parameter $\exp(\bm{Z}_a^\top\bm{\gamma})$ is 
\begin{equation}\label{eq:loglik:poisson}
l\left(\bm{\hat{\mu}}, \bm{\gamma}; \bm{w}\right) =\sum_{a \in \mathcal{A}} w_a \left(\hat{\mu}_a 
\bm{Z}^\top_a\bm{\gamma} -\exp \left(\bm{Z}^top_a\bm{\gamma}\right) -\log (\hat{\mu}_a!)\right).
\end{equation}
if the observations $(\hat{N}_a,\bm{Z}_a)_{a \in \mathcal{A}}$ are weighted by a vector  $\bm{w}$. 
Define $\bm{\gamma}=(\bm{\beta}^\top, \bm{a}^\top, \bm{b}^\top)$ with 
\[
\bm{a}=\bm{u}-\log\bm{\hat{n}}, \;\; \bm{b}=\bm{v}-\log\bm{\hat{m}}.
\]
Then with $\bm{Z}$ and $\bm{w}$ defined in Theorem~\eqref{thm:mmlogit}, we have\footnote{Note that $\bm{Z}_i$ should be interpreted here as row $i$ of the matrix  $\bm{Z}$.}
\begin{align*}
\left(\bm{Z}\bm{\gamma}\right)_{xy} & =
(\bm{\phi}_{xy}
\bm{\beta}-u_x+\log\hat{n}_x-v_y+\log\hat{m}_y)/2\\
\left(\bm{Z}\bm{\gamma}\right)_{x} &= -u_x+\log\hat{n}_x\\
\left(\bm{Z}\bm{\gamma}\right)_{y} &= -v_y+\log\hat{m}_y;
\end{align*}
and up to constant terms, $l$ and $L$ are identical.

\subsection{Proof of theorem~\ref{thm:mmlogit-var}}
The variance-covariance matrix of $\hat{\bm{\gamma}}$ follows directly from the fact that it  maximizes~(\ref{eq:loglik:poisson}), and hence is an M-estimator, see chapter~5 of~\citecomplete{vaart:1998}. The maximization of~\eqref{eq:loglik:poisson} gives  first-order conditions
\[
\sum_{a \in \mathcal A} w_a \exp(\bm{Z}_a\hat{\bm{\gamma}})\bm{Z}_a=
\sum_{a \in \mathcal A} w_a \hat{\mu}_a \bm{Z}_a,
\]
so that, applying the delta method, we get at first order
\[
\left(\sum_{a \in \mathcal A} w_a \exp(\bm{Z}_a\hat{\bm{\gamma}})\bm{Z}_a^\top\bm{Z}_a\right)(\hat{\bm{\gamma}}- \bm{\gamma}) =
\sum_{a \in \mathcal A} w_a \bm{Z}_a  (\hat{\mu}_a - \mu_a). 
\]
so we obtain a consistent estimator of the variance of $\bm{\hat{\gamma}}$ as
\[
\hat{V}\hat{\bm{\gamma}}= \bm{\hat{A}}^{-1}\; \hat{\bm{B}} \; \bm{\hat{A}}^{-1}
\]
where 
\[
\bm{\hat{A}}=\sum_{a \in \mathcal A} w_a \exp(\bm{Z}_a\hat{\bm{\gamma}})\bm{Z}_a^\top\bm{Z}_a
\]
and
\[
\bm{\hat{B}} = \sum_{a,a^\prime \in \mathcal A} w_a w_{a^\prime} \mbox{cov}(\hat{\mu}_a,\hat{\mu}_{a^\prime}) \bm{Z}^\top_a\bm{Z}_{a^\prime}.
\]

\end{appendix}

\end{document}